\documentclass[reprint,superscriptaddress,showpacs,amsmath, amssymb,aps,pra]{revtex4-2}
\usepackage{multirow}
\usepackage{amsthm}
\usepackage{amsmath}
\usepackage{latexsym}
\usepackage{amsfonts}
\usepackage{amssymb}
\usepackage{color}
\usepackage{bbm,dsfont}
\usepackage{graphicx}
\usepackage{hyperref}
\usepackage{subfigure}
\usepackage{caption}
\usepackage{xr}
\usepackage{mathrsfs}
\usepackage[noend]{algpseudocode}
\usepackage{algorithmicx,algorithm}
\usepackage{graphicx}
\usepackage{subfigure}
\usepackage{[number}
\usepackage{float}
\usepackage{appendix}
\usepackage{diagbox}

\definecolor{gray}{RGB}{123,123,123}

\newtheorem{proposition?}{Proposition?}

\theoremstyle{definition}

\newcommand{\tr}[1]{\textrm{tr}\left[#1\right]} 

\begin{document}
\title{Quantifying Quantum Steering with Limited Resources: A Semi-supervised Machine Learning Approach}
\author{Yansa Lu}
\author{Zhihua Chen}
\thanks{chenzhihua77@sina.com}
\affiliation{School of Science, Jimei University, Xiamen 361021,China}
\author{Zhihao Ma}
\thanks{mazhihao@sjtu.edu.cn}
\affiliation{School of Mathematical Sciences, MOE-LSC, Shanghai Jiao Tong University, Shanghai 200240, China}
\affiliation{Shanghai Seres Information Technology Co., Ltd, Shanghai 200040, China}
\affiliation{Shenzhen Institute for Quantum Science and Engineering, Southern University of Science and Technology, Shenzhen 518055, China}
\author{Shao-Ming Fei}
\thanks{smfei@mis.mpg.de}
\affiliation{School of Mathematical Sciences, Capital Normal University, Beijing 100048, China}
\affiliation{Max Planck Institute for Mathematics in the Sciences, 04103 Leipzig, Germany}

\begin{abstract}
Quantum steering, an intermediate quantum correlation lying between entanglement and nonlocality, has emerged as a critical quantum resource for a variety of quantum information processing tasks such as quantum key distribution and true randomness generation. The ability to detect and quantify quantum steering is crucial for these applications. Semi-definite programming (SDP) has proven to be a valuable tool to quantify quantum steering. However, the challenge lies in the fact that the optimal measurement strategy is not priori known, making it time-consuming to compute the steerable measure for any given quantum state. Furthermore, the utilization of SDP requires full information of the quantum state, necessitating quantum state tomography, which can be complex and resource-consuming. In this work, we utilize the semi-supervised self-training model to estimate the steerable weight, a pivotal measure of quantum steering. The model can be trained using a limited amount of labeled data, thus reducing the time for labeling. The features are constructed by the probabilities derived by performing three sets of projective measurements under arbitrary local unitary transformations on the target states, circumventing the need for quantum tomography. The model demonstrates robust generalization capabilities and can achieve high levels of precision with limited resources.
\end{abstract}

\maketitle

\section{Introduction}
Quantum steering, as a kind of quantum correlation lying between quantum nonlocality and quantum entanglement, holds substantial research significance and promising  applications in areas such as quantum communication and quantum computing\cite{RMPsteer, Jphysp}. It describes the ability of one party to influence the state of a distant quantum system owned by another party, through local measurements and operations on their own quantum system. The concept of quantum steering is rooted in the renowned EPR paradox\cite{Einstein} and in the pioneering work of Schr$\ddot{o}$dinger\cite{Schrodinger}. However, a rigorous definition of quantum steering was formally proposed until 2007\cite{Wiseman}. The steerable states have been proved to be a strict subset of the entangled states, and a strict superset of the states that can exhibit Bell nonlocality\cite{Jones}. Distinct from quantum entanglement and nonlocality, quantum steering is marked by its asymmetric nature. For instance, there exist quantum states $\rho_{AB}$ which are steerable from Bob to Alice, but not vice versa\cite{Bowles}. This asymmetry makes quantum steering a valuable tool for certifying the existence of entanglement between a user with an untrusted device and another user with a trusted device\cite{one-sided}. As a pivotal concept in quantum information, broad applications of quantum steering have been found in various quantum information processing tasks. These applications span the field of quantum communications and quantum computation, such as semi-device independent quantum key distribution, randomness certification, sub-channel identification and the establishment of quantum networks\cite{Branciard, Passaro, Coyle, Li, Xiang, Yueh, Wilkinson, He, Chiu, Gheorghiu, Fitzsimons, Ku, Piani, Sun, Chen, Hsieh}.

Many different kinds of methods have been proposed to detect quantum steering, such as the criteria based on linear and nonlinear inequalities, those based on uncertainty relations and moment matrix approach et. al \cite{Chen1, Bowles1, Cavalcanti, Saunders, Walborn, Pramanik, Kogias, Das, Chen2, Yang1}. Various measures have been developed to quantify quantum steering\cite{Kaur, Ku1, Skrzypczyk, Resource, Cavalcanti1}. In particular, the concepts of steerable weight and robustness have been introduced to quantify steering in any finite-dimensional bipartite quantum states\cite{Skrzypczyk}. These quantifiers can be computed through semidefinite programming (SDP) \cite{Vandenberghe, Cavalcanti1}, enabling the investigation of the quantum steering for any bipartite quantum states. However, the calculation of the steerable weight and robustness for quantum states using SDP necessitates complete knowledge of these states, typically requiring quantum tomography. Moreover, as the number of measurements performed by one party increases, the computational demand of SDP escalates, rapidly becoming impractical and time-intensive.

Learning algorithms have been widely applied in quantum computing\cite{Wise, Strikis}, quantum communication\cite{comm,qkd, Krenn}, quantum tomography\cite{Guarneri}, quantum photonics\cite{Kysheudv}, quantum metrology\cite{Huangjh}, and verifying quantum nonlocality, steering and entanglement\cite{Lu,brunner,zhaohui,Yung,Tian,Chaves,chengjie, LifengE, Yang, Changliang, Hongbin, LifengS, Kanhe, xiaoting}. In contrast to the advancements in detecting quantum correlations, only limited progresses have been made to quantify quantum correlations using machine learning\cite{Zhang, quanent}. Machine learning techniques offer a promising approach for detecting and quantifying quantum steering without the need to explore an extensive sets of measurement directions, thus outperforming SDP in terms of efficiency and speed\cite{Zhang}.
Nonetheless, to generate the feature vectors required for training models that can quantify quantum steering, complete information about the quantum states is still essential\cite{Zhang}. Moreover, to develop a highly accurate supervised machine learning model, a substantial number of labeled data points are required, which can result in time-consuming\cite{Zhang}.

Semi-supervised machine learning techniques have been used to detect quantum entanglement and steering\cite{LifengE,LifengS}. To reduce the time complexity and improve the efficiency, we employ the semi-supervised self-training (SSS) neural network model to quantify quantum steering. Additionally, measurement statistics data, obtained from measuring the involved quantum states by using local measurement devices, have been utilized as input features for neural networks to certify the genuine multipartite entanglement\cite{zhaohui}.
Drawing inspiration from this strategy, we select the feature vectors for our model as the probabilities derived from three sets of projective measurements under arbitrary local unitary transformations performed on the unknown states. This approach obviates the need for prior knowledge of the quantum states. In this work, by taking the estimation of the steerable weight as a case study, we demonstrate the effectiveness of the semi-supervised self-training model. Moreover, this method is not limited to steerable weight estimation, but can be also extended to estimate other measures of quantum correlations, provided these measures can be approximated through semi-definite programming or other mathematical techniques.

\section{Preliminary}

Quantum steering represents a distinct type of quantum correlation between two or more quantum systems in quantum mechanics. Beyond the realm of classical physics, it describes the phenomenon that a measurement of one particle immediately affects the state of the others when the multiple quantum particles are entangled, no matter how far apart they are.
Consider a bipartite quantum state $\rho_{AB}$ shared by Alice and Bob, the probability $P(a, b| x, y)$ admits local hidden variable-local hidden state model (LHV-LHS) if $P(a, b| x, y)$ satisfies the following condition,
\begin{eqnarray}
\begin{aligned}\label{eq1}
P(a, b| x, y)&=\tr{(M^{a}_{x}\otimes M^{b}_{y})\rho_{AB}}\\
&=\sum\limits_{\lambda}p(\lambda)p(a|x,\lambda)p_{Q}(b|y,\lambda),
\end{aligned}
\end{eqnarray}
where $P(a, b| x, y)$ is the probability obtained when Alice and Bob perform measurements $x$ and $y$ with measurement outcomes $a$ and $b$ on $\rho_{AB}$, denoted by $M_x^a$ and $M_y^b$, respectively, and $p_{Q}(b|y,\lambda) = \tr{M^{b}_{y}\rho_{\lambda}}$ with $\rho_{\lambda}$ being the local hidden quantum states specified by the parameter $\lambda$. If the probabilities obtained by performing all measurements $M_x^a$ and $M_y^b$ on $\rho_{AB}$ satisfies the relation (\ref{eq1}), the state $\rho_{AB}$ is said to be not steerable from Alice to Bob. Otherwise, we say that the state is steerable from Alice to Bob.

Quantum steering can be also defined through the assemblage $\{\sigma_{a|x}\}$ of Bob's unnormalized conditional states after Alice's measurement $M_x^a$.
The assemblage $\{\sigma_{a|x}\}$ admits local hidden state model (LHS) if it satisfies the following decomposition,
\begin{eqnarray}
\begin{aligned}\label{eq2}
\sigma_{a|x}=\text{tr}_A[(M^{a}_{x}\otimes I^{b})\rho_{AB}]=\sum\limits_{\lambda}p(\lambda)p(a|x,\lambda)\rho_{\lambda},
\end{aligned}
\end{eqnarray}
where $\sum\limits_{a} M_{x}^{a}= I$, $\sum\limits_{a} \sigma_{a|x}= \rho^{B}$ and $\text{tr}(\sum \limits_{a}\sigma_{a|x})=1$, $\forall a,x$. The quantum state $\rho_{AB}$ is said to be unsteerable from Alice to Bob if all the assemblages obtained by performing any local measurements by Alice admit LHS model. Otherwise, if there exists a measurement $M_x^a$ performed by Alice such that the assemblage does not admit a decomposition of (\ref{eq2}), we say that $\rho_{AB}$ is steerable from Alice to Bob \cite{Cavalcanti1}.

To quantify the quantum steering of quantum states, various quantifications including steerable weight and steerable robustness have been proposed.
An assemblage $\{\sigma_{a|x}\}$ can be
written as a convex combination of an unsteerable assemblage $\{\sigma_{a|x}^{US}\}$ and a steerable assemblage $\{\sigma_{a|x}^{S}\}$ as follows
\begin{eqnarray}
\begin{aligned}\label{eq3}
\sigma_{a|x}=\mu \sigma_{a|x}^{US}+(1-\mu)\sigma_{a|x}^{S}
\end{aligned}
\end{eqnarray}
The steerable weight of $\{\sigma_{a|x}\}$, denoted by $SW(\{\sigma_{a|x}\})$, is the minimum weight $1-\mu$ in such a decomposition with respect to all possible choices of
the unsteerable assemblage $\{\sigma_{a|x}^{US}\}$ and the steerable assemblage $\{\sigma_{a|x}^{S}\}$. The steerable weight can be calculated through SDP as follows \cite{Cavalcanti1},
\begin{eqnarray}
&&\min 1-\rm{Tr}\sum\limits_{\lambda}\sigma_{\lambda}\\ \nonumber
&& \quad\text{w.r.t.}  \quad\{\sigma_{\lambda}\}\\ \nonumber
&& \quad\quad\text{s.t.} \quad\sigma_{a|x}-\sum\limits_{\lambda}D(a|x,\lambda)\sigma_{\lambda}\geq 0\\ \nonumber
&&\quad\quad\sigma_{\lambda}\geq 0, \quad\text{for all} \quad\lambda
\end{eqnarray}
Furthermore, the steerable weight of a quantum state $\rho_{AB}$, denoted as $SW(\rho_{AB}),$
is the maximum of all $SW(\{\sigma_{a|x}\})$ over all assemblages $\{\sigma_{a|x}\},$  which are obtained by performing  any possible
measurement strategy of Alice. This process may include infinitely many measurements with arbitrary numbers of outcomes, which is time consuming as the number of measurements
becomes larger.

To address the aforementioned challenge,  machine learning approach based on artificial neural networks is adopted to estimate the steerable weight\cite{Zhang}. The lables for the training set are derived through SDP. The dataset $\mathcal{D}=\{\mathbf{x}_k,y(\mathbf{x}_k)\}_{k=1}^N$, here each $\mathbf{x}_k$ is a $16$-dimensional feature vector composing of the entries of $k$-th quantum state $\rho_k^{AB}$,  $y(\mathbf{x}_k)$ represents the steerable weight of $\rho_k^{AB}$  computed by SDP in three measurement settings with $N=170000$.
Once the model has been trained, it is capable of predicting or estimating the steerable weight for other states. However, the feature vectors of the training set are constructed by all the entries of the density matrices within the machine learning model. To estimate the steerable weight of other unknown states, state tomography is required, which remains complex and time-consuming.

\section{Semi-supervised self-training}

To improve the efficiency of machine learning, we utilize the information mined in the unlabeled data. The integration of unlabeled data in training facilitates the model's ability to discern more nuanced and expansive features and patterns, thereby bolstering its generalization capabilities. Through extensive numerical experimentation, we identify an effective approach - the semi-supervised self-training (SSS) model. In an effort to circumvent state tomography, we perform several sets of projective measurements under arbitrary local unitary operations on the target states to obtain the probabilities. These probabilities are subsequently employed as feature vectors within our SSS framework. Through numerical computation, we find that the SSS model with features obtained by performing at least three sets of projective measurements under local unitary operations can demonstrate better generalization capabilities.

\subsection{Semi-supervised Self-training Model}

The Semi-Supervised Self-Training model (SSS) is an innovative machine learning approach that integrates supervised learning algorithms with self-training techniques. This semi-supervised learning strategy leverages the power of extensive unlabeled data combined with a limited set of labeled data, broadening the horizon of data application and facilitating the extraction of more nuanced information. Furthermore, the incorporation of unlabeled data enhances the model's ability to identify more generalized features and patterns, thereby improving its performance and generalization capabilities on unseen data. Consequently, the SSS model is particularly well-suited for scenarios where there is an abundance of unlabeled data and a scarcity of labeled data, with the objective of harnessing the unlabeled data to enhance the model performance. Given the time-consuming nature of obtaining labels, specifically the steerable weight, we employ the semi-supervised self-training model for the estimation of steerable weights.

\begin{figure}[htp]
\centering
\includegraphics[width=8.6cm]{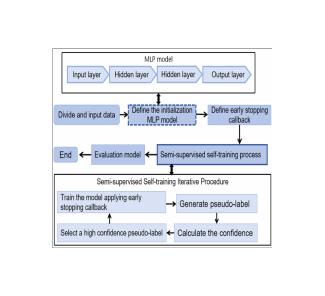}
\parbox{8.6cm}{\caption{The sketch for SSS model. The structures for the multi-layer perceptron (MLP) model is illustrated in the top box, the steps for the semi-supervised self-training is represented in the middle one, and the iterative procedure for the semi-supervised self-training model is shown in the bottom one.}\label{Figure1}}
\end{figure}

In the self-training process, the initial model is first refined by using the labeled dataset. Following this initial training process, the model is utilized to predict labels for the unlabeled data, thereby producing pseudo-labels. A confidence metric, such as the probability of a prediction or the associated uncertainty, is then applied to filter and select high-confidence pseudo-labels, which are subsequently incorporated into the training set. After this enlargement of the dataset, the model is retrained with the augmented data. The performance of the refined model is then assessed. The whole process is depicted in FIG. \ref{Figure1}.

To avoid overfitting and save computational resources in this process, we implement an early stopping callback function. This is a widely adopted regularization strategy that monitors the model's performance, halting training when the validation loss fails to decrease. This helps the model perform well without extra useless training. It also means that we do not need to set the training epochs manually.

We select the Multi-Layer Perceptron (MLP) model, a fundamental and pivotal architecture commonly referred to as an artificial neural network, for the training of our labeled data. The MLP excels in capturing non-linear relationships and is adept at managing intricate tasks. Through adjustments to the network architecture, such as the number of layers and neurons, along with the judicious selection of activation functions and training parameters, the model can be tailored to accommodate various scenarios and data types. Our MLP configuration comprises an input layer, an output layer, and two hidden layers containing 128 and 64 neurons, respectively. We employ the Rectified Linear Unit (ReLU) as our activation function and utilize the Adam optimizer for training purposes.

To assess the performance of our model, we employ two standard metrics: the mean squared error (MSE) and the coefficient of determination. These metrics are widely utilized to evaluate the performance of regression models. The MSE represents the mean of the squared differences between each predicted value and its corresponding true one. The formula for computing the MSE is provided as $MSE=\frac{1}{n}\sum_{i=1}^{n}(y_i-\hat{y}_i)^2$, where $n$ is the sample number, $y_i$ is the $i$-th true value, and $\hat{y}_i$ is the $i$-th prediction. The coefficient of determination, commonly denoted as R-squared ($R^2$), provides an indication on how well the data fitting the regression model, or how well the model predicting the outcome.  The $R^2$ value ranges from 0 to 1, $R^2=1$ indicates a perfect fit of the model to the data, whereas $R^2=0$ signifies that the model fails to offer any explanatory value. The  formula for $R^2$ is
\begin{eqnarray}
\begin{aligned}\label{eqr2}
R^2=1-\frac{\sum_{i=1}^{n}(y_i-\hat{y}_i)^2}{\sum_{i=1}^{n}(y_i-\overline{y})^2}
\end{aligned}
\end{eqnarray}
where $\overline{y}$ is the average of the true value.

\subsection{Preparing Data Set}

To facilitate the training of the semi-supervised self-training model, we must  preprocess the data firstly. The dataset employed in our study is identical to that used \cite{Zhang}. However, to circumvent the need for quantum state tomography, each feature vector in the dataset is constructed from the probabilities derived from performing several sets of projective measurements under arbitrary local unitary transformations generated randomly on the quantum states.

{We initially consider three distinct scenarios using one set of projective measurement bases: a set of entangled measurement bases, a set of separable bases, and a set consisting of two entangled vectors alongside two separable vectors.
Training the SSS model by using the features constructed from the probabilities obtained by performing these three sets of measurement bases under arbitrary local unitary matrices generated randomly,
we obtain the determination coefficients $R^2=0.126,$ $R^2=0.538$ and $R^2=0.280$, respectively. It is understood that the closer the $R^2$ to 1, the better the outcome. The performance of these three sets of projective measurements is not optimal. Given that the result from using the bases  mixing of two entangled and two separable vectors lies between the results from the fully entangled and fully separable bases, we will focus solely on either the fully entangled or the fully separable bases in the following sections.}

{Subsequently, we explore three scenarios involving two sets of projective measurements: one set of entangled bases paired with one set of separable bases, two sets of entangled bases, and two sets of separable bases, respectively. Notably, the results for the combination of one set of entangled bases and one set of separable bases are more promising, with a coefficient of determination reaching 0.747.}

{Most significantly, we investigate three scenarios of three sets of projective measurements: those consisting entirely of separable bases, those composed of one set of entangled bases and two sets of separable bases, and those made up of one set of separable bases alongside two sets of entangled bases. The coefficient of determination is about 0.824, 0.975 or 0.551, respectively.}

Through {the above} numerical computation, we determine that three sets of projective measurements are optimal, with two of these sets consisting of product bases and one set comprising an entangled basis. To show the efficiency of the SSS model, we adopte three different methods to attain the probabilities. Additionally, to show the usefulness of the local unitary operations, we chose seven sets of projective measurements without utilizing any local unitary transformations performed on the target states to generate the probabilities that form the $28$-dimensional feature vectors of the SSS model.

The feature vectors are obtained through the following measurement settings.

\begin{figure}[htp]
\centering
\includegraphics[width=8.6cm]{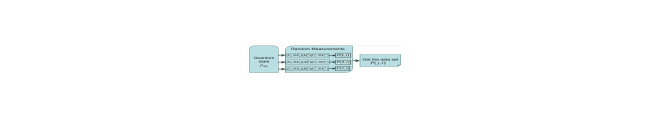}
\parbox{9cm}{\caption{The sketch to obtain F.V.-1, given by the probabilities $P(j,i)$ $(j=1,2,3,i=1,2,3,4)$ for the SSS model. The probabilities are derived by performing the first measurement setting, namely, the three sets of projective measurements subjected to arbitrary local unitary operations, on the target states.}\label{FV1}}
\end{figure}

(1) The first measurement setting: three sets of projective measurements under arbitrary local unitary transformations.
The three sets of projective measurements are
$M^j_i=|v^j_i\rangle\langle v_j^1|$ ($j=1,2,3$, $i=1,\cdots,4$) with
\begin{eqnarray}
&&|v^1_1\rangle=\frac{1}{\sqrt{2}}[1, 0, 0, 1]^{T},~ |v^1_2\rangle=\frac{1}{\sqrt{2}}[1, 0, 0, -1]^{T}\\ \nonumber
&&|v^1_3\rangle=\frac{1}{\sqrt{2}}[0, 1, 1, 0]^{T},~ |v^1_4\rangle=\frac{1}{\sqrt{2}}[0, 1, -1, 0]^{T},\\ \nonumber
&&|v^2_1\rangle=|00\rangle,~ |v^2_2\rangle=|01\rangle,~ |v^2_3\rangle=|10\rangle,~ |v^2_4\rangle=|11\rangle,\\ \nonumber
&&|v^3_1\rangle=|+0\rangle,~ |v^3_2\rangle=|+1\rangle,~ |v^3_3\rangle=|-0\rangle, ~ |v^3_4\rangle=|-1\rangle,\nonumber
\end{eqnarray}
where $|+\rangle=\frac{1}{\sqrt{2}}(|0\rangle+|1\rangle)$ and $|-\rangle=\frac{1}{\sqrt{2}}(|0\rangle-|1\rangle).$
Then we have $P(j,i)=\tr{(U_{j}\otimes V_{j})(M^j_i)(U_{j}^{+}\otimes V_{j}^{+}) \rho_k^{AB}}$ $(1\leq j\leq 3, 1\leq i\leq 4),$
where  $U_{j}$ and $V_{j}$ are arbitrary  $2\times 2$ unitary matrices {which are generated randomly}, and `+' represents conjugate and transposition. The feature vectors constructed by $P(j,i)$ are then $12$-dimensional vectors. The feature vectors derived by performing the first measurement setting on the target quantum states in the dataset are denoted as F.V.-1 for convenience, as shown in FIG.\ref{FV1}.

(2) The second measurement setting: three sets of projective measurements under local unitary operations which are selected randomly from twenty fixed ones.
The three sets of projective measurements are the same as the ones in the first measurement, but the local unitary transformations are not arbitrary. We first generate
twenty local unitary matrices randomly, then randomly choose six of these matrices to serve as
$U_j$ and $V_j$ for each target quantum state $\rho_k^{AB}$ in the dataset. The feature vectors obtained by performing the second measurement setting on the target quantum states in the dataset are denoted as F.V.-2 for convenience.

(3) The third measurement setting: seven sets of projective measurements that do not incorporate any local unitary matrices. These measurements are $M^j_i$ $(1\leq j\leq 7,i=1,\cdots,4)$, which are the same as those in the first measurement setting when $j=1,2,3$.
For $j=4,\cdots,7,$ we have
\begin{eqnarray}
&&|v^4_1\rangle=\frac{1}{\sqrt{2}}[1, 0, 0, i]^{T},~ |v^4_2\rangle=\frac{1}{\sqrt{2}}[1, 0, 0, -i]^{T}, \\ \nonumber
&&|v^4_3\rangle=\frac{1}{\sqrt{2}}[0, 1, i, 0]^{T},~ |v^4_4\rangle=\frac{1}{\sqrt{2}}[0, 1, -i, 0]^{T},\\ \nonumber
&&|v^5_1\rangle=|0H\rangle, ~|v^5_2\rangle=|0V\rangle,~ |v^5_3\rangle=|1H\rangle, ~ |v^5_4\rangle=|1V\rangle,\\ \nonumber
&&|v^6_1\rangle=|H0\rangle, ~|v^6_1\rangle=|H1\rangle,~ |v^6_3\rangle=|V0\rangle, ~ |v^6_4\rangle=|V1\rangle,\\ \nonumber
&&|v^7_1\rangle=|0+\rangle,~ |v^7_2\rangle=|0-\rangle, ~|v^7_3\rangle=|1+\rangle,~ |v^7_4\rangle=|1-\rangle,
\end{eqnarray}
where $|H\rangle=\frac{1}{\sqrt{2}}(|0\rangle+i|1\rangle)$ and $|V\rangle=\frac{1}{\sqrt{2}}(|0\rangle-i|1\rangle).$
The probabilities  $P(j,i)=\tr{M^j_i \rho^{AB}}$ $(j=1,2,\cdots,7,i=1,\cdots,4)$ and the feature vectors in the data set are all $28$-dimensional vectors. These feature vectors obtained by performing seven sets of projective measurements on the target states in the dataset are denoted as F.V.-3.

(4) The fourth measurement setting: three sets of projective measurements which are all product bases under arbitrary local unitary matrices. To examine the impact of entangled projective measurements within the SSS model, we substitute the first set of entangled projective measurements with product projective measurements $|v^7_i\rangle$, $i=1,\cdots,4$, while the second and third sets of projective measurements remain in consistent with those specified in the first measurement setting $|v^2_i\rangle$ and $|v^3_i\rangle$, respectively. The feature vectors obtained by performing the fourth measurement setting are denoted as F.V.-4.

After we attain the data set, in order to ensure the independence of the test set in the sense that it has not been used by the model in the training process, we divide the total data set which comprises $170000$ data into training and test subset. $80\%$ of the data are randomly selected to form the training data set, while the remaining $20\%$ are used as the test data set. Subsequently, to implement the SSS approach, the training dataset is further divided into unlabeled and labeled data. To show the effectiveness of the SSS model, we randomly assign $80\%$, $60\%$, $40\%$, $30\%$, and $25\%$ of the training dataset as the labeled data, with the remainders serving as the unlabeled data. This translates to $64\%$, $48\%$, $32\%$, $24\%$ and $20\%$ of the entire $170,000$ data being labeled respectively. This reduction in the amount of the labeled data not only decreases the costs associated with labeling but also mitigates the risk of overfitting, thereby enhancing the overall efficiency of data utilization.

\section{Result}

Throughout the training of the SSS model, the knowledge extracted from the labeled data is leveraged, while the unlabeled data is incorporated through self-training mechanisms. This dual approach is designed to enhance the model's generalization capabilities when encountering previously unseen data. We establish the number of iterations for the semi-supervised self-training to be 20. Initially, the labeled training data is employed to refine the model (MLP). Subsequently, the trained model is applied to predict labels for the unlabeled data, thereby generating pseudo-labels. The confidence level of these predictions is then assessed, with only the pseudo-labels exceeding a predefined confidence threshold being selected. These high-confidence pseudo-labels, along with their associated features, are amalgamated into the labeled training set for the subsequent training iteration. Following the completion of all iterations, the augmented training set is utilized to make predictions on the test set.

\begin{figure}[htp]
\centering
\resizebox{8cm}{3.6cm}{\includegraphics[width=8.6cm]{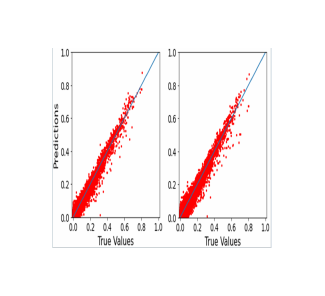}}
\parbox{8.6cm}{\caption{Comparison  the steerable weights  predicted by the SSS model and those calculated via SDP for the states within the test dataset. The horizontal (vertical) coordinate of each red point represents the steerable weight obtained by using SDP (SSS model).
The vertical coordinates of the points in the left (right) graph illustrate the results obtained by the SSS model using  F.V.-1 (F.V.-3) as the features. }\label{Figure2}}
\end{figure}

The steerable weights of the test dataset, as predicted by the trained SSS model and computed via SDP, are depicted in FIG. \ref{Figure2}. Each red dot signifies a pairing of predicted and true steerable weight values, while the blue line serves as a benchmark of equivalence, indicating where the predicted values of the steerable weight matches the true ones. As illustrated in the figure, the data points are densely clustered around the blue line, which represents equivalence between the steerable weights predicted by the SSS model and those calculated by SDP. This clustering strongly suggests that our SSS model possesses excellent generalization capabilities for handling unseen data. Consequently, the proximity of the data points to the blue line is a direct indicator of the prediction accuracy, the closer the red dots are to the blue line, the higher accuracy of the predictions is.

Due to the random selection of the training set from the total data, the random initialization or other factors that introduce randomness or variability into the training and prediction processes, learning models may give rise to slightly different results each time. We take the commonly occurring inferior results after multiple times of training, which correspond to a lower coefficient of determination $R^2$ and a higher MSE. We list these results for different feature vectors and different number of data in the training sets in Table \ref{Result1}. Compared with the high accuracy of $R^2=0.96$ in \cite{Zhang}, our model presents more effective prediction ability. To demonstrate the efficiency of the semi-supervised model, we intentionally reduced the proportion of labeled data in the training set to assess the model's performance. It is observed that $R^2$ of predictions on the test data by the model trained using just $20\%$ of the total data as labeled data remains nearly identical to 0.96 in \cite{Zhang}, except for that obtained from the model trained using the feature vectors derived by F.V.-2. This result underscores the SSS model's significant enhancement in data utilization. The MSE and $R^2$ on the test dataset predicted by the SSS model with F.V.-1 and F.V.-2 are very close to those by the model with F.V.-3, indicating that using local unitary transformations can reduce the number of projective measurements while attaining the similar predictive accuracy.
\begin{table}[htbp]
\centering{
\begin{tabular}{|cc|l|l|l|}
\hline
\multicolumn{2}{|c|}{\diagbox{portion} {F.V.}}  & F.V.-1 & F.V.-2 & F.V.-3 \\ \hline
\multicolumn{1}{|c|}{\textbf{$64\%$}}& $R^{2}$ &0.975 &0.964  &0.975\\ \cline{2-5}
\multicolumn{1}{|c|}{\textbf{}}& MSE & $1.5 \times 10^{-4}$ &  $2.2 \times 10^{-4}$&$1.5 \times 10^{-4}$\\   \hline
\multicolumn{1}{|c|}{\textbf{$48\%$}} & $R^{2}$ &0.971 &0.962 &0.973 \\   \cline{2-5}
\multicolumn{1}{|c|}{\textbf{}}& MSE & $1.74 \times 10^{-4}$ &  $2.4 \times 10^{-4}$ &$1.66 \times 10^{-4}$\\   \hline
\multicolumn{1}{|c|}{\textbf{$32\%$}} & $R^{2}$ &0.969 &0.96&0.97\\   \cline{2-5}
\multicolumn{1}{|c|}{\textbf{}}& MSE & $2.0 \times 10^{-4}$ &  $2.5 \times 10^{-4}$ &$1.83 \times 10^{-4}$\\   \hline
\multicolumn{1}{|c|}{\textbf{$24\%$}} & $R^{2}$ &0.961 &0.96 &0.965 \\   \cline{2-5}
\multicolumn{1}{|c|}{\textbf{}}& MSE & $2.4 \times 10^{-4}$ &  $2.5 \times 10^{-4}$&$2.2 \times 10^{-4}$ \\   \hline
\multicolumn{1}{|c|}{\textbf{$20\%$}} & $R^{2}$ &0.96 &0.956 &0.96 \\   \cline{2-5}
\multicolumn{1}{|c|}{\textbf{}}& MSE & $2.5 \times 10^{-4}$ &  $2.75 \times 10^{-4}$ &$2.5 \times 10^{-4}$\\   \hline
\end{tabular}}
\caption{MSE and $R^2$ on the test dataset predicted by the SSS model with different features and different portion of the total data as the labeled data. The results in the last three columns refer to the ones obtained by the model with different F.V.-i ($i=1,2,3$) shown in section III. B.}
\label{Result1}
\end{table}

In addition to the feature vectors obtained through the three measurements detailed in Table \ref{Result1}, we also consider the SSS model with F.V.-4. Utilizing
$64\%$ of the total data as the labeled data for model training, we attain $R^2=0.821$ and
$MSE=0.0014$. These findings indicate that the set of entangled projective measurements used to construct the features in SSS model plays the critical role in estimating the steerable weight.

\begin{figure}[h]
\centering
\resizebox{8cm}{8.1cm}{\includegraphics[width=8.6cm]{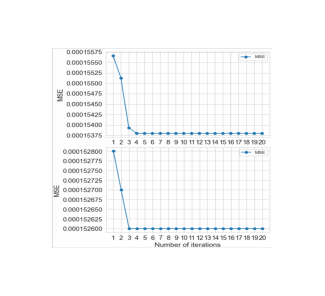}}
\parbox{8.6cm}{\caption{MSE varies with the number of iterations. The figure above (below) illustrates the variation of MSE with the number of iterations for the steerable weight within the test dataset predicted by the model with features F.V.-1 (F.V.-3).}\label{MSE2}}
\end{figure}

\begin{figure}[h]
\centering
\resizebox{8cm}{8.1cm}{\includegraphics[width=8.6cm]{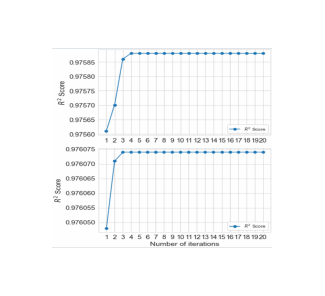}}
\parbox{8.6cm}{\caption{$R^2$ changes with the number of iterations. The above (below) figure refers to the case that $R^2$ varies with the number of iterations for the steerable weight within the test dataset predicted by the model with the features F.V.-1 (F.V.-3). It shows that $R^2$ is greater than $0.97$ in the eighth iteration, indicating that our model has effective prediction power.}\label{R2}}
\end{figure}

\begin{figure}[!htb]
\centering
\resizebox{7cm}{7.2cm}{\includegraphics[width=8.6cm]{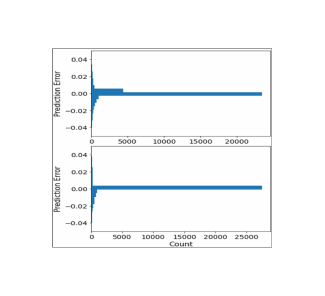}}
\parbox{8.6cm}{\caption{Distributions of the predictive errors on the test dataset. The  above (below) figure refers to the errors on the test data predicted by the model with the features F.V.-1 (F.V.-3). Most of the errors are around zero.}\label{Error}}
\end{figure}

Throughout the training process of the model, with the number of self-training iterations fixed at 20, both the MSE and the $R^2$ coefficient are subject to the change with the number of iterations progresses. Taking the dataset with features F.V.-1, where $64\%$ of the total data is used as labeled data, as an example, we illustrate the trends of MSE and $R^2$ under the evolution of each iteration, as depicted in FIG. \ref{MSE2} and FIG. \ref{R2}, respectively. Similar patterns are observed for data obtained through other measurement settings. Moreover, the error distributions on the test set are shown in FIG. \ref{Error}. It reveals that the error distributions are tightly clustered around zero, indicating a very narrow spread. This clustering suggests that the error variance is minimal, which in turn signifies that the SSS model is adept at predicting unseen data with a high level of accuracy.

\begin{figure}[!htb]
\centering
\resizebox{8.4cm}{3.7cm}{\includegraphics[width=8.6cm]{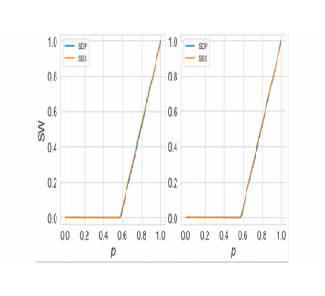}}
\parbox{8.6cm}{\caption{The steerable weight with respect to $p$ of Werner states predicted by SSS and the ones computed by SDP. The vertical coordinate `SW' represents the steerable weight. The results predicted by the SSS model using features F.V.-1 (F.V.-3) are represented by the orange line in the left (right) graph, and the ones calculated by the SDP are represented by the blue line. }\label{werner}}
\end{figure}

To show the capabilities of our model more effectively, we estimate the steerable weight of the Werner state $\rho^w_{AB}$,
\begin{eqnarray}
\begin{aligned}\label{eqw}
\rho^w_{AB}=p|W\rangle\langle W|+(1-p)\frac{I_4}{4},
\end{aligned}
\end{eqnarray}
where $|W\rangle=(|01\rangle-|10\rangle)/\sqrt{2}$, $I_4$ is the $4\times 4$ identity matrix and $p\in[0,1]$. The maximum steerable weight of $\rho^w_{AB}$ is obtained when $p=1$ and the quantum steering exists when $p > 1/\sqrt{3}$\cite{Skrzypczyk}. The total dataset comprises 300 samples. The proportion of labeled data and the subset designated for training remain consistent with those used for the 170,000 randomly generated states. Similarly, the measurement settings employed for feature extraction are identical across both datasets. The $R^2$ for the test data is approximately 0.999 predicted by the SSS models with F.V.-i for $i=1,2,3$, signifying that the SSS model exhibits exceptional generalization capabilities for unseen data. This score surpasses 0.99 given in \cite{Zhang}. As depicted in FIG. \ref{werner}, the line predicted by the SSS model closely aligns with the line derived from the SDP calculations, and it appears to be even smoother.

{The initial step to obtain experimentally the entries of a density matrix is to select a suite of observables for the reconstruction process. For 2-qubit systems, nine observables are needed for tomography of the quantum state. These observables are then measured by repeatedly preparing the same 2-qubit state and recording the expectation values. Each observable is associated with a distinct measurement configuration, and the system is subjected to each of these configurations. We gather the probability distribution of the measurement outcomes for each observable. Given that we are working with a 2-qubit system, each measurement yields one of four possible outcomes with respect to the observable's eigenstates. Utilizing the probability distributions from the 9 observables, we employ computational methods such as maximum likelihood estimation to determine the density matrix. In summary, the tomography process requires nine sets of 4-dimensional probability distributions and subsequent mathematical processing, such as maximum likelihood estimation, to extract the quantum state's entries.

In contrast, the use of a 28-dimensional feature vector has its advantages. In reality, a 12-dimensional feature vector is sufficient to achieve superior results for training. This is accomplished by performing three sets of projective measurements under random local unitary matrices on the quantum states, which yields the 12-dimensional feature vectors. In practice, the exact knowledge of the local unitary matrices is not required as they can be generated randomly. We only need to collect the probability distributions from these three sets of projective measurements. Then, this 12-dimensional vector is input into the well-trained SSS model, from which the steerable weight is derived. This approach significantly reduces the resources needed, compared with the state tomography and the semidefinite programming (SDP).}

\section{Conclusion}
We have exploited the semi-supervised self-training machine learning approach to estimate the steerable weight of unknown quantum states. The feature vectors are constructed from the probabilities obtained by performing three sets of projective measurements under local unitary transformations on the quantum states. This method significantly reduces the computational complexity, obviating the need for state tomography. Additionally, the SSS model, trained on a limited amount of labeled data, demonstrates robust predictive capabilities, thereby substantially reducing the time required for data labeling. Our model is also adaptable for estimating the different measures for other quantum correlations, provided that the corresponding measures of limited amount of states can be computed through some mathematical methods. Our study could uncover novel insights into the potential applications of quantum steering in quantum communication and computation tasks.

\

\

\

{\bf {ACKNOWLEDGEMENTS}}
We acknowledge enlightening discussions with Qiongyi He and Yu Xiang and we thank Ye-Qi Zhang for sharing the data for machine learning. This work is supported by the Fundamental Research Funds for the Central Universities; the National Natural Science Foundation of China (NSFC) under Grants  12071179, 12371132, 12075159, 12171044; the specific research fund of the Innovation Platform for Academicians of Hainan Province.

\appendix

\end{document}